\documentclass[structabstract]{aa}  
\usepackage{graphicx}
\begin{document}
%
   \title{Ubiquitous quiet-Sun jets}

   \author{V. Mart\'\i nez Pillet,
          \inst{1}
          J.C. del Toro Iniesta\inst{2}
          \and
	  C. Quintero Noda\inst{1,3}
          }

\institute{Instituto de Astrof\'\i sica de Canarias, E-38200, La Laguna, Tenerife, Spain~~~
          \email{vmp@iac.es, cqn@iac.es}
 \and
 Instituto de Astrof\'\i sica de Andaluc\'\i a (CSIC), Apdo. de Correos 3004, E-18080, Granada, Spain~~~
          \email{jti@iaa.es}
 \and
Departamento de Astrof\'\i sica, Univ. de La Laguna, La Laguna, Tenerife, E-38205, Spain~~~
            }

   \date{Received ; accepted }

 
  \abstract
  {
  IMaX/{\sc Sunrise} has recently reported the temporal evolution of highly dynamic and
  strongly Doppler shifted Stokes $V$ signals in the quiet Sun.
  }
   {
   We attempt to identify the same quiet-Sun jets in the {\em Hinode} 
   {\rm spectropolarimeter (SP) data set}.
   }
   { 
   {\rm We generate combinations of linear polarization magnetograms with blue- and redshifted 
   far-wing circular polarization magnetograms to allow an easy identification of the 
   quiet-Sun jets.}
   }
   {
   {\rm The jets are identified in the} {\em Hinode} {\rm data where both red-
   and blueshifted cases are often found in pairs. They appear next to regions of
   transverse fields that exhibit quiet-Sun neutral lines.
   They also have a clear tendency to occur in the outer 
   boundary of the granules. These regions always display highly displaced and 
   anomalous Stokes $V$ profiles}.
   }
   {
   {\rm The quiet Sun is pervaded with jets formed when new field regions emerge at 
   granular scales loaded with 
   horizontal field lines that interact with their surroundings. This interaction
   is suggestive of some form of reconnection of the involved field lines that 
   generates the observed high speed flows.}
   }


   \maketitle

\section{Introduction}

The magnetism of the quiet Sun has recently come under deep scrutiny. Thanks to
the advent of two major observatory missions, the {\em Hinode} satellite (see
Kosugi et al. \cite{Kosugi07}, Tsuneta et al. \cite{Tsuneta08}, Suematsu et al.
\cite{Suematsu08}, Shimizu et al. \cite{Shimizu08a}, Ichimoto et al. \cite{Ichimoto08}) 
and the {\sc Sunrise} balloon (see Barthol et al. \cite{Barthol11} and Solanki et al.
\cite{Solanki10}), a wealth of information about the physical processes behind
the magnetism of the quiet Sun has been gathered. In particular, the existence
of emerging flux in the form of granular scale loops is now solidly
established. From Lites et al. (2008; see also Orozco et al. 2007), who
published the most detailed map of linear polarization signals marking the
presence of transverse fields, to \cite{Mariam08} who analysed a large number
of internetwork loop appearances, we have formed a somewhat detailed picture
of how new flux arrives at the solar surface on scales as small as one
arcsecond. (For a larger statistical sample of flux emergence cases with detected 
horizontal fields, we
refer to Ishikawa \& Tsuneta \cite{Ishikawa09} and Jin et al. \cite{Jin09}.)
Emerging flux is observed first in the form of upflowing horizontal
internetwork features (HIFs as first coined by \cite{Lites96}), which, at a
later stage, displays footpoints of opposite polarities at both ends.
\cite{Ishikawa10} describes with {\em Hinode} the rise of a single, more or
less coherent, flux concentration through the line-forming region of the
spectropolarimeter (SP) instrument.  The {\sc Sunrise} mission has described in
detail similar processes (that even resolve substructures within the emerging
region, see \cite{Danilovic10a}) and produced a statistically sound set of
properties and evolutionary timescales of the HIFs (\cite{Danilovic10b}).

   \begin{figure*}
   \centering
   \includegraphics[width=20cm,angle=0]{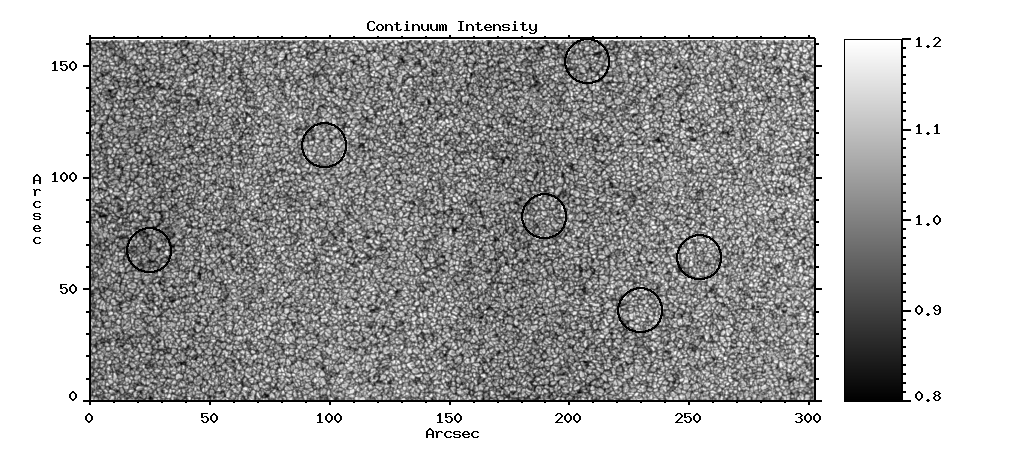}
   \includegraphics[width=20cm,angle=0]{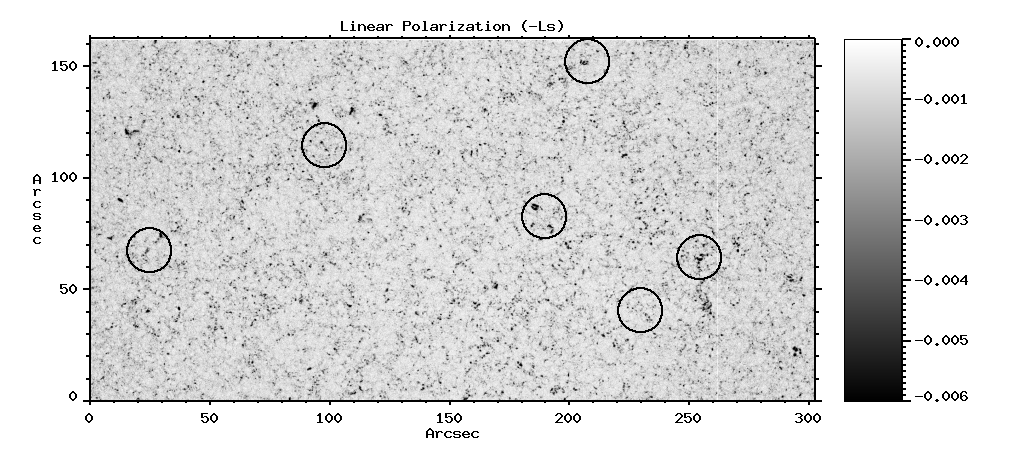}
   \caption{Top: {\em Hinode}/SP continuum map constructed from the individual slit
   scans. Bottom: Fe {\sc i} 6302.5 \AA~$L_s$ magnetogram (actually $-L_s$
   for later comparison) constructed from Eq.\  
   \ref{equ:samples}. The circles mark the regions shown in Fig.\ \ref{Combo}.
              }
              \label{FigSP1}%
    \end{figure*}

The complexity of these flux emergence processes must be somehow reflected in
the observed Stokes profiles of the quiet Sun. In particular, the circularly
polarized Stokes $V$ profiles are known to display very asymmetric shapes from
the Advanced Stokes Polarimeter (ASP) era (Sigwarth \cite{Sigwarth01}) to
the more recent {\em Hinode}/SP observations (\cite{Viticchie11}). The origin
of these anomalously behaved profiles is unclear and the physical processes
involved are largely unknown. Single lobed and/or multilobed shapes often show
large displacements from the central (rest) wavelengths, which are indicative of
large line-of-sight (LOS) flow velocities.  \cite{Shimizu08b} found strong
(supersonic) downflows, in, both, the vicinity of sunspots and in
the quiet Sun. These downflows were interpreted as representing the formation of a
strong field concentration through the process of convective collapse (see also
Nagata et al. \cite{Nagata08} and Fischer et al.
\cite{Fischer09}). One of the few other cases where the origin of
these profiles has been  associated with a physical scenario was presented by
Socas-Navarro \& Manso Sainz (\cite{Navarro05}). These authors found quiet-Sun
Stokes $V$ profiles with a third lobe, this time blueshifted by 7.5 km s$^{-1}$
from the rest position. They tentatively ascribed this lobe to an upward
rebound of the material participating in the convective collapse process.
Independently of the validity of these explanations, the
identification of the physical scenario in which these profiles are generated
is clearly crucial to developing diagnostic tools that retrieve information from them.
This is the main objective of this work.

The IMaX instrument (Mart\'\i nez Pillet et al.  \cite{Martinezpillet10})
aboard {\sc Sunrise} has provided the temporal evolution of circular
polarization signatures observed in the continuum reference point (Borrero et
al. \cite{Borrero10}; BMS hereafter).  These signals were identified to occur
close to upflowing material in the form of granules that carry with them
considerable amounts of fairly inclined fields.  Most of these events occur
near a magnetic neutral-line configuration, suggesting 
that there is a link to reconnection
processes taking place in regions of flux emergence.  High speed flows are
needed to explain the generation of significant signals at the
observed continuum wavelength. This phenomenon seems to be fundamental enough
to occur more than 400 times across a field of view (FOV) of 100 arcseconds squared
in one hour (BMS, 1.3$\times$10$^{-5}$ events/s arscec$^2$).  Understanding the
nature of these processes, the magnetic topology in which they are originated,
and quantifying the energetics involved would help us to identify 
these processes in data sets with wider spectral coverage such as those
obtained by {\em Hinode}/SP, which offers full Stokes profiles
of two magnetically sensitive Fe~{\sc i} lines. This paper provides the
conceptual bridge between the IMaX/{\sc Sunrise} observations of quiet-Sun jets
and the {\em Hinode}/SP data.

   \begin{figure*}
   \centering
   \includegraphics[width=20cm,angle=0]{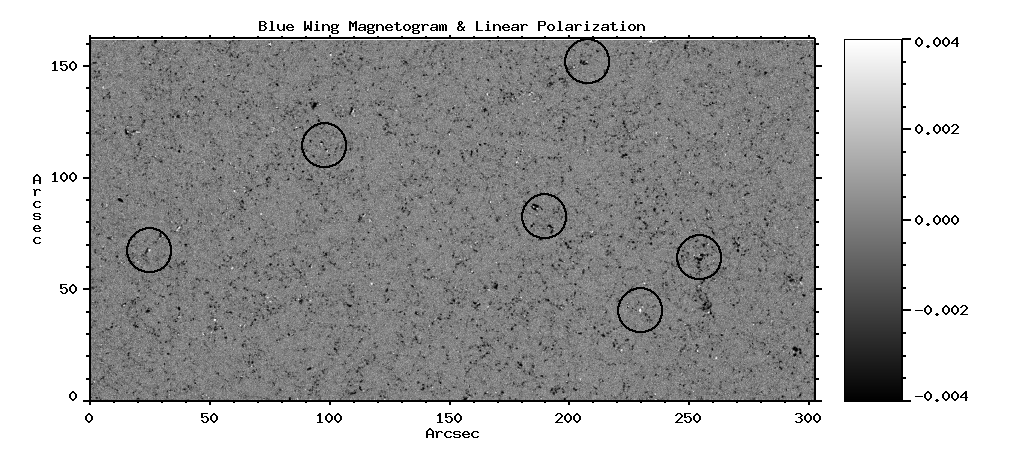}
   \includegraphics[width=20cm,angle=0]{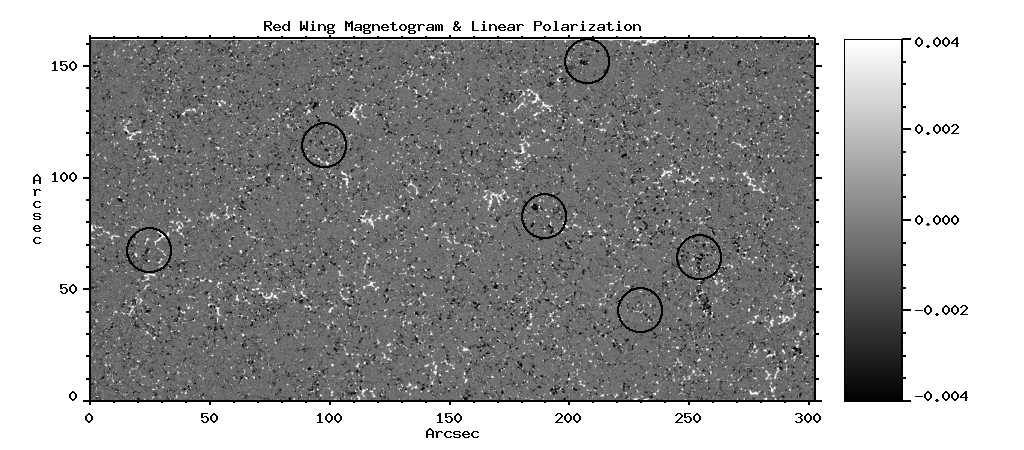}
   \caption{Top: $|V^-_c|-L_s$ magnetogram. Black signals are linear polarization and
   white signals blue-wing circular polarization.
   Bottom: same as above but for the red-wing magnetogram, $|V^+_c|-L_s$. 
   The circles mark the regions shown in Fig.\ref{Combo}.
               }
              \label{FigSP2}%
    \end{figure*}
%


\section{{\em Hinode/SP} data and comparison with IMaX/SUNRISE observations}

Before describing the data used in this work, we discuss the
observations of the IMaX instrument used by BMS. They correspond to data
sets that use all four Stokes parameters at four wavelength points within the
spectral line of IMaX (the Fe {\sc i} 5250.2 \AA~ line) and one in the
continuum. The spectral line samples were nominally taken at -80, -40, 40, and
80 m\AA~ from the central wavelength. In addition,   
we routinely observe a fifth point in the continuum at 227
m\AA~redwards of the spectral line.  
This wavelength point is half-way between the nominal IMaX line
and the Fe{\sc i} 5250.6 \AA~ line. The exact observing mode used by IMaX is
described in detail by Mart\'{\i}nez Pillet et al. (2010), where it is referred
to as the V5-6 observing mode (vector mode with five wavelength samples and six
accumulations achieving a signal-to-noise ratio, hereafter $S/N$,
of 1000 per wavelength).  The spatial
resolution of these first-flight IMaX magnetograms was slightly better than 0.2
arcsec.

Averaging the samples within the line, one produces mean linear polarization 
($L_s$) and mean circular polarization (i.e., normal) magnetograms ($V_s$) 
\begin{equation}
\label{equ:samples}
V_s={{1}\over{4I_c}}\sum_{i=1}^4 a_i V_i ~ ~\mbox{and}~~
L_s={{1}\over{4I_c}}\sum_{i=1}^4 \sqrt{Q_i^2+U_i^2},
\end{equation}
where $\bar a$ is the vector $[1,1,-1,-1]$ and $I_c$ the mean observed
continuum intensity. Since four wavelength points are included in these
magnetograms, their $S/N$ is typically around 2000.  The circular polarization
observed at the continuum sample will hereafter be referred to as $V_c$ (and
given in units of $I_c$).  BMS found that 70 \% of the $V_c$ signals observed
by IMaX occur near strong blue shifts as measured by a Gaussian fit to the
Stokes $I$ samples. This indicates that a large fraction of these signatures
are probably generated by a blueshifted circular polarization signal from the
Fe~{\sc i} 5250.6 \AA. However downflows that generate a redshift of the nominal
IMaX line cannot be excluded.  The estimated LOS flow speeds involved were
larger than 3 km s$^{-1}$. These strong flows (most likely supersonic) were
found to occur immediately next to HIF patches observed in the $L_s$
magnetograms (see the on-line material of BMS). While all circular polarization
continuum signals (tracing the high speed flows) were associated with HIF
patches, the reverse was not true. Many HIF regions that had just appeared 
displayed no
clear signatures at the IMaX continuum point. The association identified by 
BMS appears to represent flux emergence on granular scales that interacts with
pre-existing flux and that, in some cases, reconnects with it giving rise to the
observed jets. However, with as little as four points within the line and one in
the continuum on only one side of the spectral line, it is hard to
investigate these quiet-Sun jets. Identifying this process in the {\em
Hinode}/SP database will help us enormously in understanding its
underlying physical mechanism.

To proceed with this identification in the {\em Hinode}/SP data set, we use the
same observations described in \cite{Lites08}. These data have a $S/N$ similar
to that of the IMaX individual wavelength points ($\approx$1000). The spatial
resolution is around 0.3 arcsec.  Each slit position has an exposure time of 
4.8 seconds. To produce observables comparable to those obtained by
IMaX, the {\em Hinode}/SP data was first convolved with a 95 m\AA~wide
Gaussian profile simulating the spectral resolving power of the {\sc Sunrise}
magnetograph. This number corresponds to the intrinsic resolution of the
IMaX instrument (85 m\AA) scaled by the factor $\lambda_{6302}/\lambda_{5250}$
after subtracting (quadratically) the spectral resolution of Hinode/SP (see
Tsuneta et al. \cite{Tsuneta08}).
The resulting profiles were then sampled at wavelengths similar to
those used by IMaX scaled, again, by the same wavelengths ratio (in
particular, the continuum points are now located at $\pm$272 m\AA~from line centre).
The line selected for this study is the Fe {\sc i} 6302.5 \AA~line as it has a
very similar Zeeman sensitivity as the nominal IMaX line. This time, we also
include a continuum sample on the red side of the line to help us 
identify up- and downflow events. These $V^{+-}_c$ continuum magnetograms ($+$
and $-$ superscripts correspond to the red and blue samples, respectively) are
reminiscent of those used by Ichimoto et al. (2009) in their study of the
penumbral flows. The $L_s$ and $V^{+-}_c$ magnetograms can now be combined in a
way similar to that demonstrated by BMS to identify locations where 
the HIF concentrations are
associated with nearby high-speed flows.

Figure \ref{FigSP1} shows the continuum (top) and $-L_s$ magnetogram (bottom)
from the {\em Hinode}/SP set that are similar to those published by Lites et
al. (2008; cf. their Figs. 1 \& 2). They are given here for easy cross
reference with that work. The black circles indicate the regions 
explored in detail in the next section. The top panel of Fig.~\ref{FigSP2}
displays the combined $|V_c^-|-L_s$ magnetogram. Strong unsigned-blue-wing
continuum signals are shown in white, while strong linear polarization signals
are displayed in black.  The number of identified coincidences of HIF features with
nearby blueshifted signatures is enormous: 209 by simple visual inspection
(thresholds for this identification were 0.32 \% in $|V_c^-|$ and 0.28 \% in
$L_s$). {\em This large number of identifications indicates that the phenomenon
described by BMS is also present in the {\em Hinode}/SP data}. In 24 cases
(10 \% of the total), we found a clear blueshifted pattern without any
evident linear polarization signals in the surroundings. Given the almost
ubiquitousness of the association between these two observables, we hypothesize
that the linear polarization signals existed at a time other than the exact time when
the slit sampled that region (or that the signals remain below the noise
level).  This hypothesis is reinforced from the visualization of the on-line
material of BMS, which illustrates how time-dependent these events are.  

The rather different nature of these instruments makes it difficult to compare
their occurrence rates.  However, a rough comparison can be made if one notes
that at any given time in the on-line material of BMS there are in the range of 5-10
events over a FOV of 45$\times$45 arcsec$^2$. As the typical lifetime of
the supersonic jets (100 seconds, see BMS) is much longer than the exposure
time of any slit position, we can take the image in Fig.~\ref{FigSP2} as a
snapshot similar to those in IMaX (with, indeed, similar exposure times). Thus,
simple scaling of the Hinode/SP FOV (300$\times$160 arcsec$^2$) suggests that
we should find about 24 times more events, i.e., between 120 and 240 events.
These numbers compare rather well with the number of detected events here.

In the bottom panel of Fig.~\ref{FigSP2} we present the same $-L_s$ magnetogram as
before but this time combined with the red $|V^+_c|$ continuum magnetogram
(that is, the displayed quantity is $|V_c^+|-L_s$). The appearance is rather
different from the previous case as clear residuals from the network
regions are now evident. This is indeed expected as it is well known that the
Stokes $V$ profiles of this line display an extended red wing that is absent from
the blue side of the line (see, e.g., Mart\'\i nez Pillet et al.
\cite{Martinezpillet97}). These signals from the extended wings that mark
network locations complicate the identification of events similar to those
described for the blue wing magnetogram.  Although one can envisage techniques
to remove this leakage from network regions, in this work we simply use
the magnetogram as presented in this figure as it still allows this
identification to be made (see next section). 

From a quantitative point of view, we note that the most
easily identifiable white events (the high speed flows) in the $|V^{+-}_c|$
magnetograms of Fig. \ref{FigSP2} are in the range from 0.01 to 0.03 after
convolving with the IMaX spectral function. IMaX signals have a similar 
magnitude before the image reconstruction process and reach values two times 
larger after reconstruction (with a resolution of 0.15 arcsec).

\section{Examples of quiet-Sun jets identified in {\em Hinode}/SP}

\begin{figure*}
   \centering
    \includegraphics[width=23cm,angle=90]{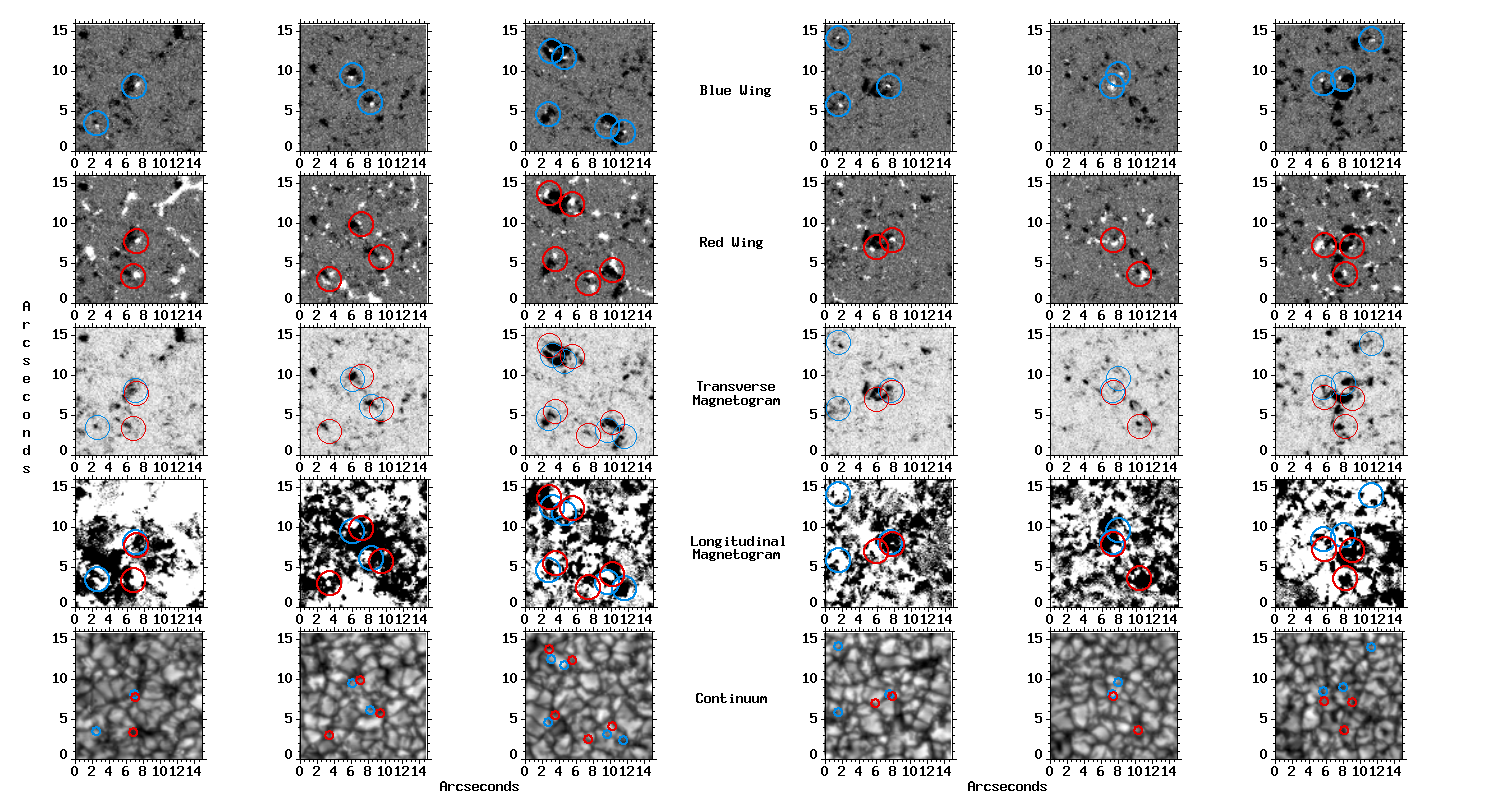}
      \caption{Six selected cases of quiet-Sun jets. The order from left to right is
      the same as that shown in Figs. \ref{FigSP1} and \ref{FigSP2} by the black
      circles.  First row: $|V^-_c|-L_{s}$ (scaled to $\pm
      0.004$).  The blue circles mark the identified coincidences between HIFs
      and the high-speed flows. Second row: $|V^+_c|-L_{s}$
      (same scaling). The red circles mark the identified coincidences between
      HIFs and the high-speed flows. Third row: $-L_s$ from Eq.\ \ref{equ:samples} 
      (circles are thinner for easier visualization of the transverse field signals).
      Fourth row: Normal $V_s$ magnetogram scaled to $\pm 0.001$. Fifth row:
      Associated continuum frames. The blue and red circles 
      have been reduced in size to help us identify the locations
      within the granules where the high-speed flows occur. 
      }
         \label{Combo}
   \end{figure*}

From a visual inspection of the $|V^-_c|$ and $L_s$ combination, we 
selected six cases for closer analysis. They are marked in Figs. \ref{FigSP1} and
\ref{FigSP2} with black circles and shown in an expanded view in Fig. \ref{Combo}.
The first row displays the combination $|V^-_c|-L_s$
The second row shows the same regions, but now using $|V^+_c|$. The
third row is the transverse magnetogram $-L_s$ (same as Fig. \ref{FigSP1} bottom
image). In all examples, the selected cases display a number ($>$1) of
coincidences between HIF regions and blueshifted $V^-_c$ signals (first row).
In the first row, these coincidences are marked with the blue circles 
(of 2.76 arcsec diameter).
Visual inspection of the Stokes $V$ profiles associated with the white patches
showed that they correspond to a blueshifted third lobe or to just one
isolated Doppler-shifted lobe.  The example shown in the fifth position shows a
strong blueshifted patch of circular polarization that has no clearly
identifiable linear polarization nearby. 
However, signatures of transverse
fields can be found in the corresponding panel (third row), albeit weak. 
The $|V^+_c|$ magnetograms were not used in the identification of the selected 
regions because of
the network leakage effect described before. The most clear example of this leakage
is seen in the first example: a narrow tilted zone of about 5 arcsec appears in
the top right corner of the image. (The network origin of this signal is
confirmed in the normal magnetogram presented in the fourth row.) The red
circles mark redshifted jets that are next to the corresponding HIF. In this
case, care was taken to ensure that none of the red circles marked
correspond to a network-leaked signal. This time, all Stokes $V$ profiles
displayed a third lobe displaced to the red or a strongly broadened red lobe
with a redshifted zero-crossing point. 

We note the rather common coincidence between blueshifted and
redshifted patches. They seem to occur in pairs.  While sometimes only one of
the Doppler-shifted signatures appear, often both are present next to the
corresponding HIF and next to each other. This can be checked in the 
fourth row of Fig. \ref{Combo} that presents the normal magnetogram (highly
saturated to $\pm 10^{-3}$) with overlaid blue and red circles from the
previous examples. Apart from the proximity between blue- and redshifted
patches from the quiet-Sun jets, this panel also proves that they always occur
in regions containing mixed polarities (a fundamental ingredient of any
type of reconnection). 
The neutral lines observed inside the circles might be caused by two
different reasons. First, an emerging bipole will always have its intrinsic
neutral line (on top of the transverse fields). Second, a newly emerged bipole can
encounter pre-existing opposite polarity fields nearby. That in most of
the circles seen in the fourth row of Fig. \ref{Combo} one can identify several
neutral lines suggests that both types are present. However, it is not yet clear
whether the jets are related to either one of these type of neutral lines or to
both.  This point will be addressed in future work.

Finally, the fifth row provides the corresponding continuum frames. The
overlaid circles have been made smaller in this case (0.92 arcsec diameter) to
help us identify the exact location of the high-speed jets on top of the granular
pattern. As can be easily inferred, these signatures occur near the edges of
the granules in the vast majority of the selected cases. This location suggests
a link with the fields that would be encountered by the emerged bipole
in the surrounding intergranular lanes.

\section{Conclusions}

The quiet-Sun jets discovered by BMS have been unambiguously identified in the
{\em Hinode}/SP data, proving that this is an ever present process that pervades
the solar surface. Using the same combination of linear polarization and Stokes
$V$ far-wing magnetograms as used in the IMaX analysis, 
we have developed a natural way
of pinpointing them in the {\em Hinode}/SP database. In this case, we find coincidences
between HIF patches and strongly Doppler-shifted signals in both the red and
the blue sides of the spectrum. While no analysis of the profiles is presented in
this paper, it has been verified that they correspond to cases where the Stokes $V$
signal displays highly asymmetric shapes with multiple lobes far from the reference
rest wavelengths. The profiles are  similar to those presented in Socas-Navarro
and Manso Sainz (\cite{Navarro05}) in their Fig. 2 and a deeper study of the 
{\em Hinode}/SP data will likely deduce velocities similar to those
found in that work. 

The results presented here, that is, the existence of blue- and redshifted
flows near HIFs (but not on top of them), the presence of nearby neutral 
lines in normal magnetograms (i.e. mixed polarities),
and the location of the jets on the granular boundaries provide a 
consistent picture
of the physical process behind these quiet-Sun jets. Granulation emerges to the
surface loaded with magnetic fields forming a tangled loop configuration; the
horizontal fields emerging with the granules encounter nearby pre-existing
fields at the intergranular regions and, sometimes, reconnect with them. This
naturally explains the existence of the high speed flows, the mixed polarity
character, and their spatial location within the granules and at some distance
from the transverse field patch. Sweet-Parker reconnection,
as modeled by Chae et al. (\cite{Chae02}) and Litvinenko et al. (\cite{Litvinenko07}),
and favored by the low electrical conductivity of the photospheric layers,
can be invoked to explain the observed flows. According to these works, the 
outflowing speeds from the reconnection region will be in the range of 3-10 km s$^{-1}$
(see their Figure 6 of the first reference and Equation 17 of the second) which are
capable of generating the observed jets.

Two other possible physical mechanisms could be considered to explain
these jets. One is a combination of convective collapse processes 
and their associated downflows together with occasional rebounded material 
propagating upwards (see Socas-Navarro and Manso Sainz \cite{Navarro05} or
Nagata et al. \cite{Nagata08}). The other mechanism is siphon flows along 
magnetic arches (Montesinos \& Thomas \cite{Montesinos93}). While all three 
possibilities need to be studied carefully, the strong association with 
flux emergence favors the reconnection scenario.
The {\em Hinode}/SP
database now provides a fantastic opportunity to study in more detail (with
Stokes inversion techniques) the topology and the energetics of these
ubiquitous quiet-Sun jets identified now by, both, IMaX/SUNRSIE and {\em
Hinode}/SP. We urgently need to understand which counterparts of these events 
can be seen in different wavelengths and in higher altitudes in the solar
atmosphere.

%
%

\begin{acknowledgements}
Comments by J. M. Sykora and by J.M. Borrero are gratefully acknowledged.  The
work has been funded by the Spanish MICINN under project AYA2009-14105-C06-03/06 and
by Junta de Andaluc\'{\i}a under project P07-TEP-02687. {\em Hinode} is a
Japanese mission developed and launched by ISAS/JAXA, with NAOJ as domestic
partner and NASA and STFC (UK) as international partners. It is operated by
these agencies in co-operation with ESA and NSC (Norway).
\end{acknowledgements}

\end{document}